\documentclass{desyproc}

\begin{document}
\title{Extending axions searches with a spherical TPC}

\author{{\slshape Javier~Galan$^1$, Gilles Gerbier$^{2}$, Ioannis~Giomataris$^{2}$, Thomas~Papaevangelou$^2$, Ilias~Savvidis$^3$}\\[1ex]
$^1$University of Zaragoza (Spain)\\
$^2$CEA Saclay\\
$^3$University of Thessaloniki (Greece)}

\contribID{galan\_javier}

\desyproc{DESY-PROC-2013-XX}
\acronym{Patras 2013} 
\doi  

\maketitle

\begin{abstract}
We present the prospects for detection of KK-axions using a large volume spherical TPC through natural decay to two gammas. The higher excited mass states of this axion model allows to reach densities which could be detectable by this method. We show the capability of this detector to detect 2-prong events coming from rest-mass axion decays and we provide efficiencies obtained under some gas mixtures and pressure conditions. The sensitivity limit of a future experiment with existing detectors geometry has been estimated for the case of zero background limit.
\end{abstract}

\section{Introduction}

The axion is a hypothetical neutral pseudoscalar particle which was already predicted in 1977. This new weakly interacting particle came out as an elegant solution to the CP problem of strong interactions in QCD~\cite{Peccei}. We focus on an extension to this model. In superstring theories it turns out to be possible to lower the string scale without lowering the Planck scale~\cite{Arkani,Dienes}. One of the most interesting features of the higher-dimensional axionic theories is that their mass spectrum consists of a tower of Kaluza-Klein (KK) excitations, which have an almost equidistant mass-space related to the compactification radius R~\cite{Raffelt}. Moreover, the coupling to photons, $g_{a\gamma}$, of the excited states is of the same order as the coupling of the ground state, which is naturally identified with the QCD axion.

\vspace{0.2cm}




As for the case of standard QCD axions, a considerable range of KK-axion mass modes (up to about 20keV) should be produced in the Sun~\cite{Raffelt}. An immediate consequence of the production of higher axion masses inside the Sun is the capability to produce low momentum axions. A small fraction of the axions produced would have a momentum which is lower than the Sun escape velocity, and these axions would keep gravitationally trapped describing elliptic orbits around the Sun~\cite{Zioutas}. The accumulation of these axions during the Sun lifetime would allow to reach detectable densities in the Earth's neighborhood (see section~\ref{sc:sens}).

\section{Spherical TPC detector}

\begin{figure}[ht]
\centerline{\includegraphics[width=0.35\textwidth]{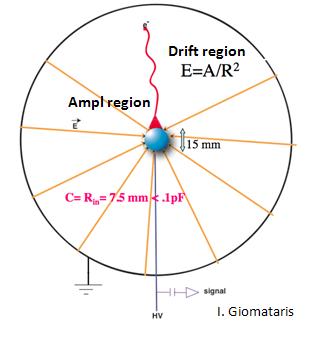}}
\caption{An schematic of the spherical TPC concept and detection principle.}\label{Fig:sph}
\end{figure}

We propose to use a spherical TPC~\cite{Giomataris} to detect KK-axions through its natural decay to two photons, by studying the topology of 2-prong events. This detector consists of a spherical grounded cavity which is filled with gas. In the center of this cavity is placed a small spherical sensor (made of metallic or resistive materials) where a high positive voltage is applied. The field produced inside the cavity allows drifting the electrons and ions produced by ionizing interactions in the gas. The field close to the sensor (typically 1\,cm diameter) is high enough to produce signal amplification through electron avalanche processes (see Figure~\ref{Fig:sph}).

The main advantages of using a spherical TPC reside on its simplicity, single read-out, large volume capability, good energy resolution (11\% FWHM at 6keV) and low electronic noise related to the low capacitance of the spherical geometry. The low energy threshold is only limited by the ionization energy of the gas (around 25 eV). The dynamic range of the detector can be adjusted by using different values of the amplification field to scan interactions from few eV to several MeV, as would be produced by alphas or heavy ions.

\vspace{-0.3cm}
\section{Detection efficiency to 2-prong events}

\begin{figure}[ht]
\centerline{\includegraphics[width=0.45\textwidth]{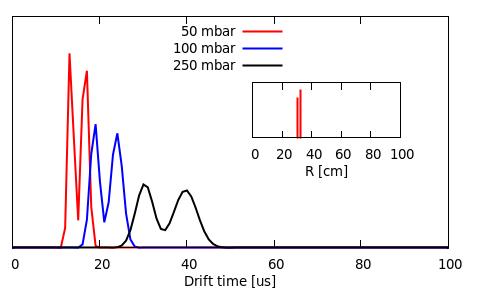}\includegraphics[width=0.45\textwidth]{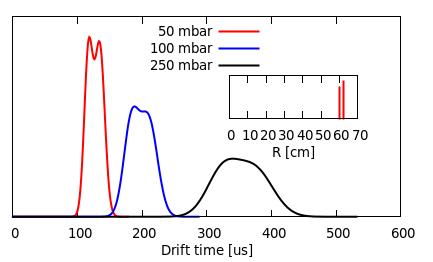}}
\caption{Effect of diffusion for a 2-prong like event at different gas pressures. Inset plots show the distance of the simulated events to the sensor, (left) two close events at about 30\,cm, (right) at about 60cm. The different pressures simulated at Argon+10\%CH4 show how relevant is the pressure choice in the capability to distinguish such events.}\label{Fig:diff}
\end{figure}

One of the key features of our detector is the capability to detect 2-prong events. In a gaseous detector there are two physical parameters that affect the probability to distinguish two gammas, the attenuation length and the charge diffusion. This two parameters depend on the gas mixture and pressure, providing the detector with certain flexibility to set the efficiency to this kind of events. The field defined by the spherical geometry produces a strong dependence on drift velocity and diffusion as a function of the distance to the sensor (see Figure~\ref{Fig:diff}).

\begin{figure}[ht]
\centerline{\includegraphics[width=0.45\textwidth]{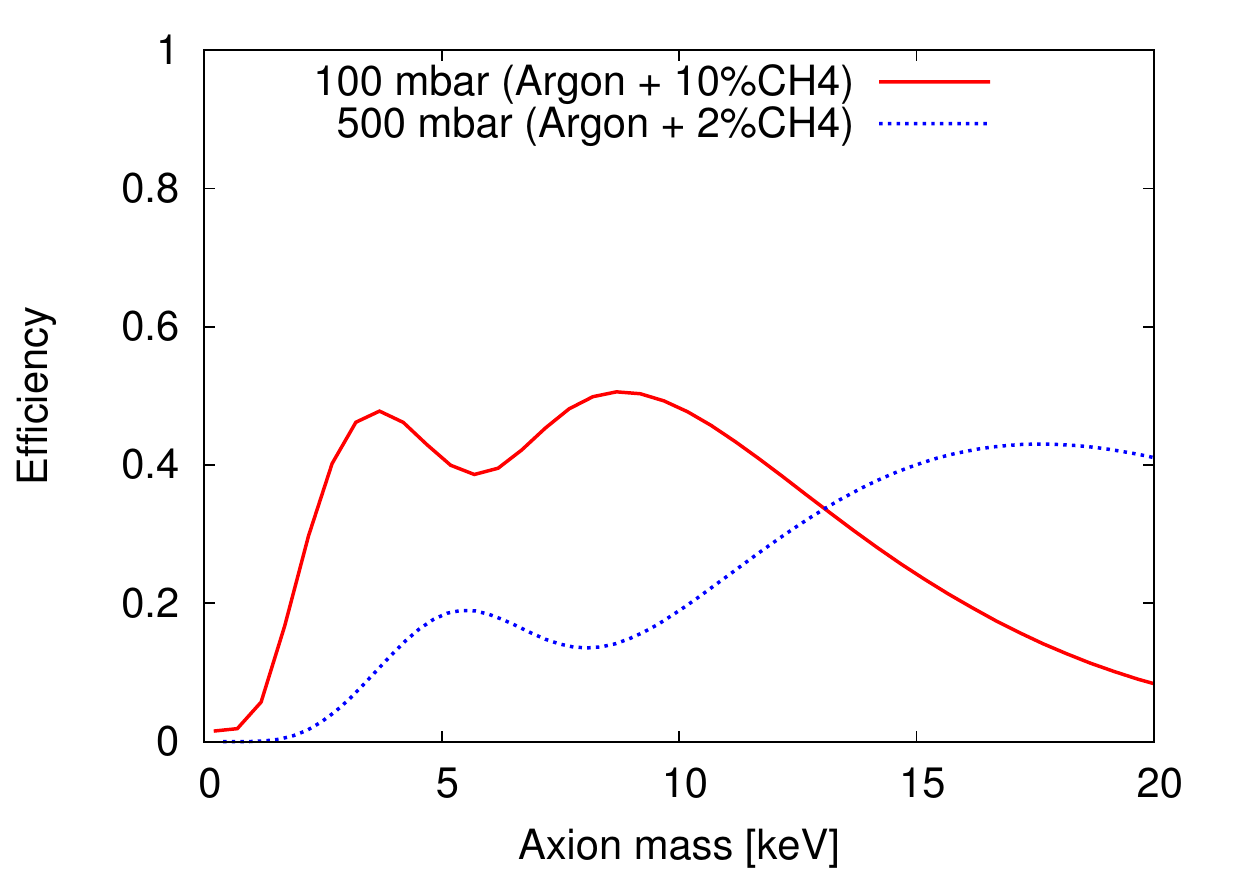}\includegraphics[width=0.45\textwidth]{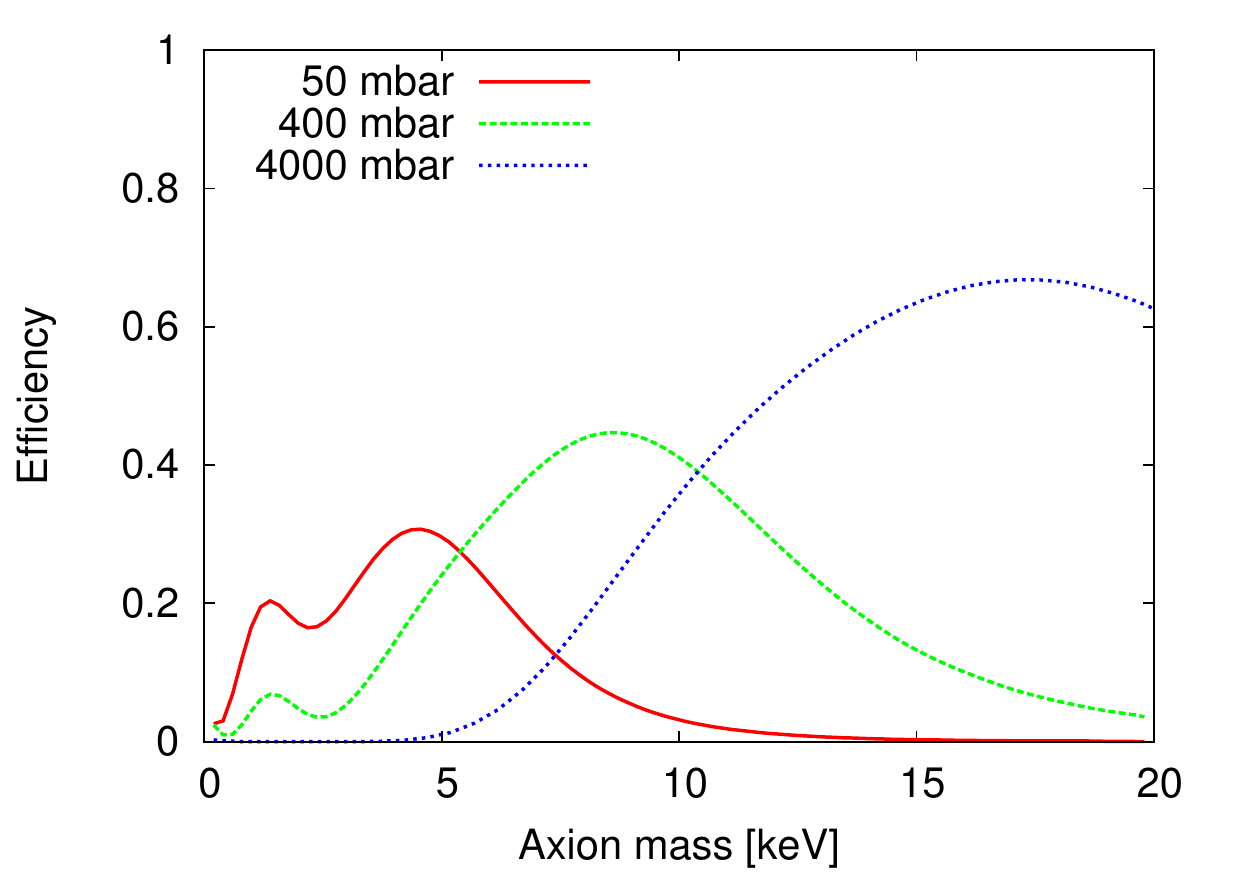}}
\caption{2-prong events detection efficiency as a function of the axion mass. (left) Gas argon+CH$_4$ and spherical detector radius 65\,cm. (right) Gas neon+2\%CH$_4$ and spherical detector radius 30\,cm.}\label{Fig:Eff}
\end{figure}

In one side, we expect the attenuation length to be as short as possible so that both gammas interact inside the detector, but in the other side, we want that the gammas to be far enough so that they can be differentiated in the detector read-out. In order to evaluate the capability to detect this kind of events in our detector we have produced Montecarlo simulations of rest-mass decays to two gammas at different pressures and gas mixtures. We present efficiencies for two typical mixtures used; argon+2\%CH$_4$, argon+10\%CH$_4$ and neon+2\%CH$_4$ (see Figure~\ref{Fig:Eff}).

It must be noticed the great flexibility on detection efficiency by using different gas mixtures and pressure, allowing to optimize the detector for this search. Different axion mass ranges can be covered with different efficiencies by varying the pressure, supporting a hypothetical positive signal which could be modulated with the detector pressure.

\section{Sensitivity prospects to KK-axions}
\label{sc:sens}

The sensitivity of our experiment for detection of decaying KK-axions will depend on the model described in~\cite{Zioutas} through the accumulated axion density ($\rho_a$) and the decay rate, or mean life ($\tau_a$). These quantities depend on the coupling of axion to photon ($g_{a\gamma}$) and the mass of the axion ($m_a$). Considering the mean life of these particles is few orders of magnitude above the age of the solar system we can express these quantities as follows

\begin{equation}
\rho_a = 1.18 \times 10^{39} \left( \frac{g_{a\gamma}}{\mbox{GeV}^{-1}} \right)^2 \left[ \mbox{m}^{-3} \right] \quad \quad \tau_a = 1.35 \times 10^5 \left( \frac{g_{a\gamma}}{\mbox{GeV}^{-1}} \right)^{-2} \left( \frac{m_a}{\mbox{eV}}^{-3} \right) \left[ \mbox{s} \right].
\end{equation}

Our sensitivity will be directly related to the total number of decays observed in a certain amount of time ($t_{exp}$) in the volume defined by our detector ($V_{sph}$) and the detection efficiency of 2-prong events ($\epsilon_{det}$). The number of decays observed is then given by the following expression,
 
\begin{equation}
N_{\gamma} = \tau_a^{-1} \cdot \rho_a \cdot V_{sph} \cdot t_{exp} \cdot \epsilon_{det}
\end{equation}
\noindent which depends finally on the value of $g_{a\gamma}$ and $m_a$.


\vspace{0.2cm}

\begin{figure}[ht]
\centerline{\includegraphics[width=0.45\textwidth]{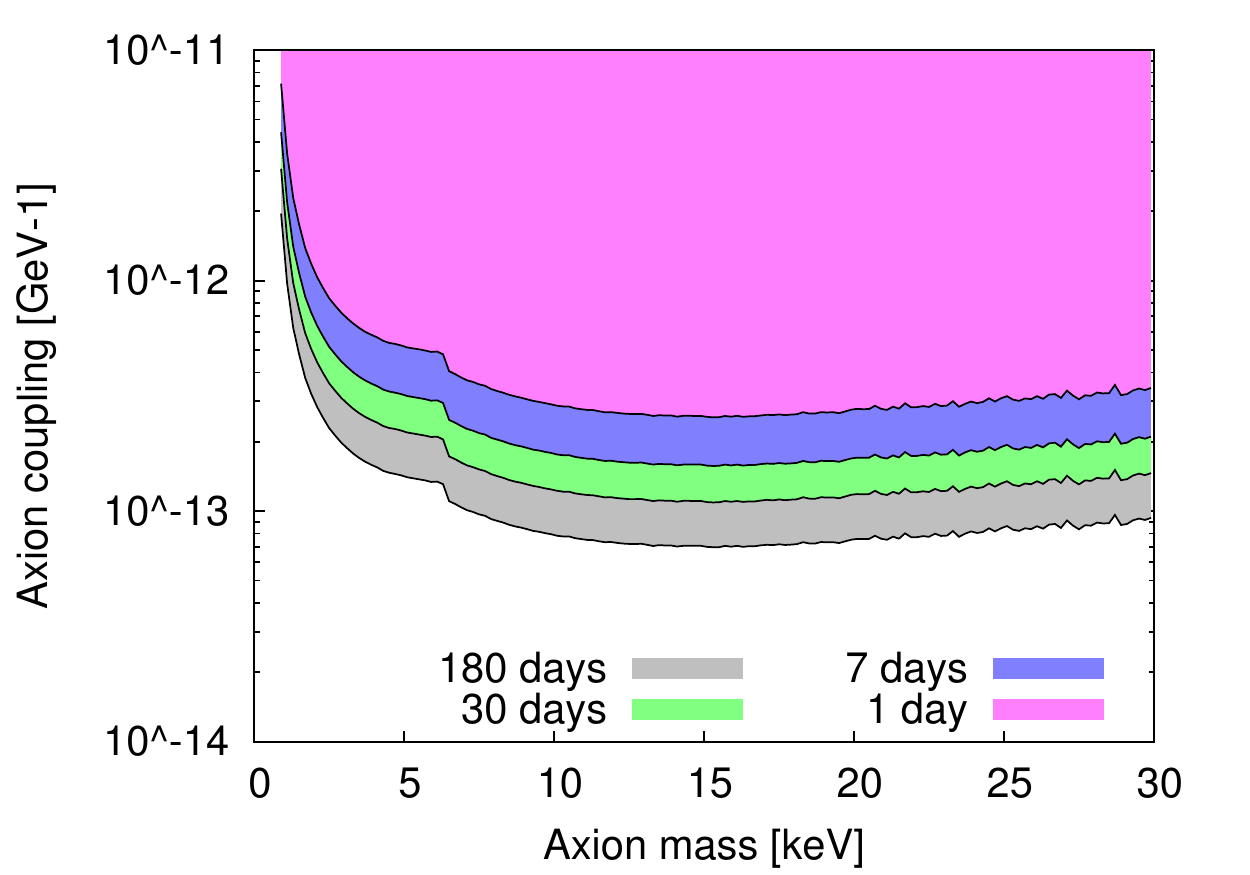}\includegraphics[width=0.45\textwidth]{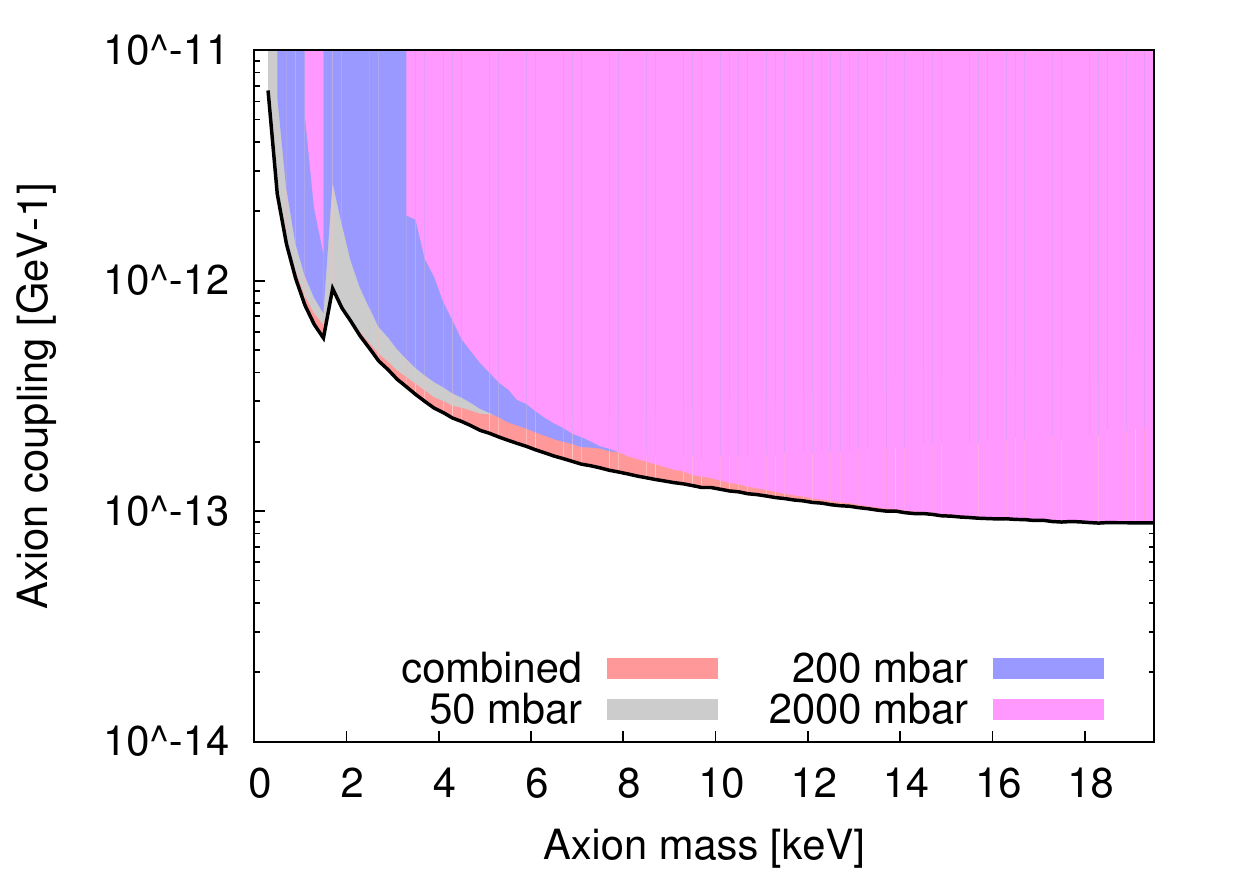}}
\caption{Sensitivity limit estimation for the axion-photon coupling in case of zero background. (left) Coupling limit for different exposure times (sphere radius 65\,cm) in an argon+10\%CH$_4$ at 100\,mbar. (right) Coupling limit for neon+2\%CH$_4$ gas mixture (sphere radius 30\,cm) at different pressures for a 180 days exposure period and the combined result.}
\label{Fig:sens}
\end{figure}

In case no signal is observed we can obtain a limit on the coupling of the axion to photon. For simplicity, here we present the sensitivity limit of such experiment in the case of zero background which provides a rough estimate of the limits achievable (see Figure~\ref{Fig:sens}). Further work on this direction should include the measured background level of the detector after necessary optimization.

\section{Discussion}
We have shown the sensitivities reachable with a spherical TPC for KK-axions search. The existence of KK-axion implies the existence of the QCD axion, thus the estimated sensitivities for this search correspond to an unexplored region on the QCD axion parameter space. We conclude then that the sensitivities reachable with this type of experiment are competitive enough to prove the existence of KK-axions (and thus QCD axions). In case no signal is found we could set the first experimental limit on this type of axions.


\begin{footnotesize}

\end{footnotesize}


\end{document}